\documentclass[twocolumn,showpacs,preprintnumbers,amsmath,amssymb]{revtex4}
\usepackage{graphicx}
\usepackage{dcolumn}
\usepackage{bm}

\begin{document}

\preprint{}

\title{{\it In situ} photoemission study on atomically-controlled La$_{1-x}$Sr$_x$MnO$_3$ thin films: Composition dependence of the electronic structure}

\author{K. Horiba}
\altaffiliation[Present address: ]{RIKEN/SPring-8, 1-1-1 Kouto, Mikazuki-cho, Sayo-gun, Hyogo 679-5148, Japan}
\author{A. Chikamatsu}
\author{H. Kumigashira}
\email[Author to whom correspondence should be addressed; Electronic mail: ]{kumigashira@sr.t.u-tokyo.ac.jp}
\author{M. Oshima}
\affiliation{Department of Applied Chemistry, The University of Tokyo, 7-3-1 Hongo, Bunkyo-ku, Tokyo 113-8656, Japan}

\author{N. Nakagawa}
\author{M. Lippmaa}
\altaffiliation[Also at ]{Combinatorial Materials Exploration and Technology (COMET), Tsukuba 305-0044, Japan}
\affiliation{Institute for Solid State Physics, The University of Tokyo, Kashiwa 277-8581, Japan}

\author{K. Ono}
\affiliation{Institute of Materials Structure Science, High Energy Accelerator Research Organization (KEK),  1-1 Oho, Tsukuba 305-0801, Japan}

\author{M. Kawasaki}
\altaffiliation[Also at ]{Combinatorial Materials Exploration and Technology (COMET), Tsukuba 305-0044, Japan}
\affiliation{Institute for Materials Research, Tohoku University, Sendai 980-8577, Japan}

\author{H. Koinuma}
\altaffiliation[Also at ]{Combinatorial Materials Exploration and Technology (COMET), Tsukuba 305-0044, Japan}
\affiliation{Materials and Structures Laboratory, Tokyo Institute of Technology, Yokohama 226-8503, Japan}

\date{\today}

\begin{abstract}
We have investigated change in the electronic structures of atomically-controlled La$_{1-x}$Sr$_x$MnO$_3$ (LSMO) thin films as a function of hole-doping level ($x$) in terms of {\it in situ} photoemission spectroscopy (PES) and x-ray absorption spectroscopy (XAS) measurements. The {\it in situ} PES measurements on a well-ordered surface of high-quality epitaxial LSMO thin films enable us to reveal their intrinsic electronic structures, especially the structure near the Fermi level ($E_F$). We have found that overall features of valence band as well as the core levels monotonically shifted toward lower binding energy as $x$ was increased, indicating the rigid-band like behavior of underlying electronic structure of LSMO thin films. The peak nearest to $E_F$ due to the $e_g$ orbital is also found to move toward $E_F$ in a rigid-band manner, while the peak intensity decreases with increasing $x$. The loss of spectral weight with $x$ in the occupied density of states was compensated by simultaneous increment of the shoulder structure in O 1$s$ XAS spectra, suggesting the existence of a pseudogap,  that is depression in spectral weight at $E_F$, for all metallic compositions. These results indicate that the simple rigid-band model does not describe the electronic structure near $E_F$ of LSMO and that the spectral weight transfer from below to above $E_F$ across the gap dominates the spectral changes with $x$ in LSMO thin films.
\end{abstract}


\maketitle

\section{INTRODUCTION}
Hole-doped manganese oxides with perovskite structure of $Re_{1-x}Ae_x$MnO$_3$ ($Re$ and $Ae$ being trivalent rare earth and divalent alkaline earth elements, respectively), have attracted considerable attention, because they exhibit a rich phase diagram originating from collective phenomena under the competition among different electronic phase due to close interplay among spin, charge, orbital, and lattice degree of freedom \cite{MIT_Rev, CMR_Book}. Among these manganites, La$_{1-x}$Sr$_x$MnO$_3$ (LSMO), thin films of which have been investigated here, is one of the most prototypical compounds. A parent compound LaMnO$_3$ is an antiferromagnetic insulator while hole-doping induced by substitution of Sr for La in the parent compound produces a ferromagnetic metallic phase \cite{Urushibara}, which shows colossal magnetoresistance (CMR). In addition to the CMR effect, the highest Curie temperature ($T_C$) of 360 K among manganites and half metallic nature \cite{JHPark} make LSMO intriguingly attractive for potential application to magnetoelectoronic devices. Further hole-doping beyond optimal doping level for ferromagnetic states ($x$ = 0.4) induces a transition from ferromagnetic metal to antiferromagnetic metal states for $x >$ 0.5 \cite{Fujishiro}. In order to clarify the origin of these unusual physical properties, it is important to investigate how the electronic structures change as a function of hole-doping (Sr-concentration $x$). 

Photoemission spectroscopy (PES) has long played a central role in the measurement of the electronic structutes of manganese oxides and their changes with carrier doping \cite{Chainani, Saitoh1, Sarma, Saitoh2, Matsuno}. Nevertheless, recently questions arose as to reliability of PES spectra on addressing the bulk electronic structure of manganese oxides; the PES spectra of manganese oxides, especially the density of states (DOS) at the Fermi level ($E_F$), strongly depend on the surface preparation procedure as well as the experimental conditions. Since PES is a quite surface sensitive technique owing to short photoelectron mean free paths, this may originate from different surface electronic structure induced by different surface preparations (scraping, fracturing, or sputtering and annealing) as well as different surface sensitivity under different experimental conditions \cite{Sekiyama_Nature}. Although a large number of PES studies have been made on manganites, there is little agreement as to the hole-induced change in electronic structure of LSMO.  Furthermore, the metallic states of LSMO derived from the coherence of doped states may be deeply disturbed by the disorder induced by conventional surface preparations (for example, in scraped surface of polycrystalline samples) \cite{Joynt}. Thus, in order to understand the bulk electronic structure, it is indispensable to perform the PES measurements on well-defined surfaces of manganese oxides. 

In this paper, we report an {\it in situ} PES and x-ray absorption spectroscopic (XAS) study on well-ordered surfaces of LSMO thin films grown by laser molecular beam epitaxy (laser MBE). Rapid progress in high-quality crystal growth techniques using laser MBE \cite{Koinuma} has enabled us to grow LSMO thin films with an atomically flat surface. We found that the valence band spectra show significant difference between an atomically-flat surface and a scraped one; in particular, the spectral weight of Mn 3$d$ $e_g$ states closest to $E_F$ is suppressed strongly in scraped surfaces. We also found the spectra of ferromagnetic LSMO films ($x$ = 0.4) exhibits a clear evidence for a Fermi cut-off, reflecting their metallic nature. This result clearly demonstrates the importance of {\it in situ} PES study on a well-ordered surface of manganese oxides for revealing their intrinsic electronic structure. Combining the PES spectra and O 1$s$ XAS spectra on the high-quality surface of LSMO thin films with various hole-doping levels, we have successfully obtained a picture of how the electronic structure evolves from antiferromagnetic insulator to ferromagnetic metal through the observation of the chemical potential shift and spectral weight transfer near $E_F$.

\begin{figure}
\includegraphics[width=0.8\linewidth]{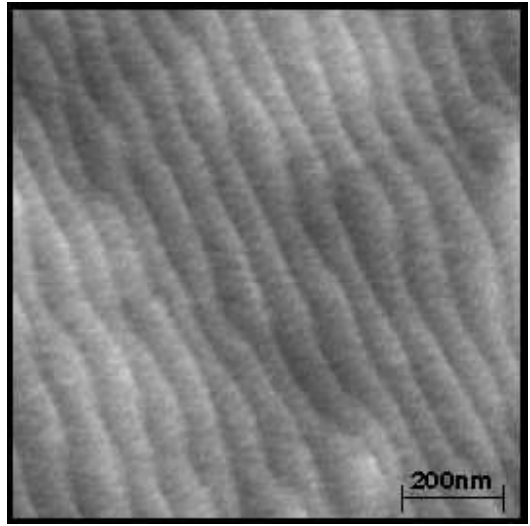}
\caption{\label{figure1}A typical AFM image of a 400 {\AA} thick La$_{0.6}$Sr$_{0.4}$MnO$_3$ film grown on a SrTiO$_3$ (001) substrate.}
\end{figure}

\section{EXPERIMENTAL} 
Experiments were carried out using the high-resolution synchrotron radiation PES system combined with a laser MBE chamber \cite{Horiba_RSI}, which was installed at beamline BL-2C of Photon Factory \cite{BL2C}. The LSMO thin films were grown epitaxially on SrTiO$_3$ (STO) substrates by pulsed laser deposition. Sintered LSMO ($x$ = 0, 0.1, 0.2, 0.3, 0.4, and 0.55) pellets were used as ablation targets. A Nd: YAG laser was used for ablation in its frequency-tripled mode ($\lambda$ = 355 nm) at a repetition rate of 0.33 Hz. The wet-etched STO (001) substrates were annealed at 1050 $^{\circ}$C at an oxygen pressure of 1 $\times$ 10$^{-6}$ Torr before deposition in order to obtain an atomically flat TiO$_2$-terminated surface \cite{Kawasaki}. LSMO thin films with a thickness of about 400 {\AA} were deposited on the TiO$_2$-terminated STO substrates at a substrate temperature of 950 $^{\circ}$C and at an oxygen pressure of 1 $\times$ 10$^{-4}$ Torr. The intensity of the specular spot in reflection high energy electron diffraction (RHEED) pattern was monitored during the deposition to determine the surface morphology and the film growth rate. Epitaxial growth of LSMO thin films on the STO substrate was confirmed by the observation of clear oscillations due to the layer-by-layer growth mode. The LSMO thin films were subsequently annealed at 400 $^{\circ}$C for 30 minutes in atmospheric pressure of oxygen to remove oxygen vacancies. After cooling to below 100 $^{\circ}$C, the films were moved into the photoemission chamber under a vacuum of 10$^{-10}$ Torr.  The in-vacuum transfer is necessary to obtain the highest quality surface as revealed by {\it in situ} - {\it ex situ} comparative PES measurements \cite{Horiba_RSI}. The PES spectra were taken using the GAMMADATA SCIENTA SES-100 electron-energy analyzer. The total energy resolution at the photon energy of 600 eV was about 150 meV. The Fermi level of the samples was referred to that of a gold foil which was in electrical contact with the sample. XAS spectra were obtained by measuring the sample drain current. The surface stoichiometry of the samples was carefully characterized by analyzing the relative intensity of relevant core levels, confirming that the composition of the samples was the same as the ceramic targets \cite{Kumi_APL}.  The surface morphology of the measured films was analyzed by {\it ex situ} atomic force microscopy (AFM) in air. The crystal structure was characterized by four-circle x-ray diffraction (4c-XRD), confirming epitaxial growth of the films on the substrates. Magnetization was measured by a superconducting quantum interference device (SQUID) magnetometer with the magnetic field applied along the [100] axis parallel to the surface. Electorical resistivity was measured by the four-probe method.

\begin{figure}
\includegraphics[width=0.8\linewidth]{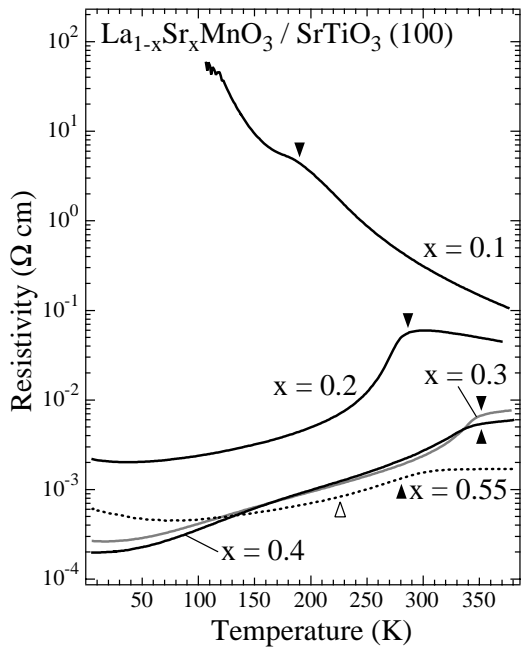}
\caption{\label{figure2}Temperature dependence of resistivity for LSMO thin films. Filled triangles indicate the critical temperature for the ferromagnetic phase transition determined by the SQUID measurements. Anomalies indicated by open triangles are due to the antiferromagntic transition.}
\end{figure}

\begin{table}
\caption{\label{table1}Physical properties of fabricated LSMO thin films. $T_C$ and the N\'eel temperature ($T_N$) are determined by the SQUID measurements. The out-of-plane
($c$-axis) and the in-plane ($a$-axis) lattice constants ($d_{\perp}$ and $d_{//}$, respectively) are determined by the 4c-XRD measurements.}
\begin{ruledtabular}
\begin{tabular}{l|ccccc}
\multicolumn{1}{c|}{$x$} & $T_C$ (K) & $T_N$ (K) & $d_{\perp}$ ({\AA}) & $d_{//}$ ({\AA}) & Ground State\\
\hline
0 & - & - & 3.949 & 3.892 & -\\
0.1 & 190 & 96 & 3.926 & 3.904 & AFI\\
0.2 & 287 & - & 3.880 & 3.905 & FM\\
0.3 & 352 & - & 3.855 & 3.905 & FM\\
0.4 & 354 & - & 3.830 & 3.904 & FM\\
0.55 & 281 & 226 & 3.803 & 3.889 & AFM\\
\end{tabular}
\end{ruledtabular}
\end{table}

\section{RESULTS AND DISCUSSION}
\subsection{Physical properties of the LSMO thin films}
Figure~\ref{figure1} shows a typical AFM image of the fabricated LSMO $x$ = 0.4 thin film. Atomically-flat step-and-terrace structures which reflect the surface of STO substrates are clearly observed. The step height is measured to be about 0.4 nm which is close to the $c$-axis constant of the LSMO films, indicating that film surfaces can be controlled on an atomic scale. Such atomically-flat step-and-terrace structures are also observed in all the films reported here with different compositions, indicating the successful control of surface structure. Figure~\ref{figure2} shows temperature dependence of electrical resistivity ($\rho$) for LSMO thin films with different compositions. Filled triangles indicate $T_C$ determined by magnetization measurement, which nearly agree with those of the kink in the $\rho-T$ curve. A significant change in $\rho$ is observed at around $T_C$. Above $x$ = 0.2, metallic conduction is observed in the low-temperature ferromagnetic phase ($T < T_C$), whereas $\rho$ at $x$ = 0.1 exhibits insulating behavior in all temperature range. Note that the $\rho-T$ curve for $x$ = 0 films, as well as that for $x$ = 0.1 below 100 K could not be measured owing to the limitation of our $\rho-T$ apparatus. In the high-temperature paramagnetic phase ($T > T_C$), the $\rho-T$ curve is still characteristic of nonmetal (semiconductor), i.e., $d\rho/dT <$ 0, for $x$ = 0.2, the curve becomes metallic with further increasing $x >$ 0.3. Anomalies indicated by open triangles for $x$ = 0.55 are due to the magnetic transitions from ferromagnetic to antiferromagnetic phase, respectively. We summarize in table~\ref{table1} the physical properties of the LSMO films. The obtained values are in good agreement with published data \cite{Izumi_APL, Izumi_PRB, Fukumura}.

\begin{figure}
\includegraphics[width=0.95\linewidth]{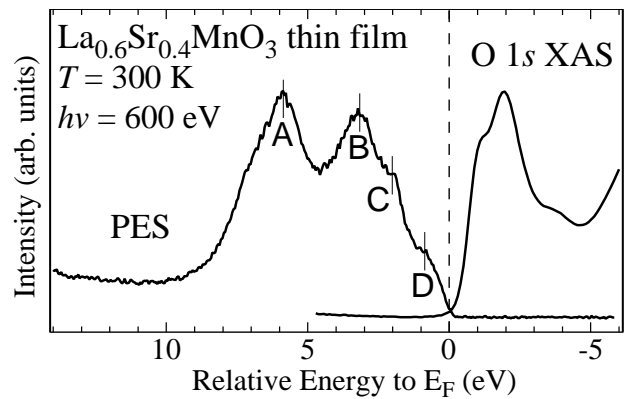}
\caption{\label{figure3}PES and XAS spectra of La$_{0.6}$Sr$_{0.4}$MnO$_3$ thin films.}
\end{figure}

\begin{figure}
\includegraphics[width=0.95\linewidth]{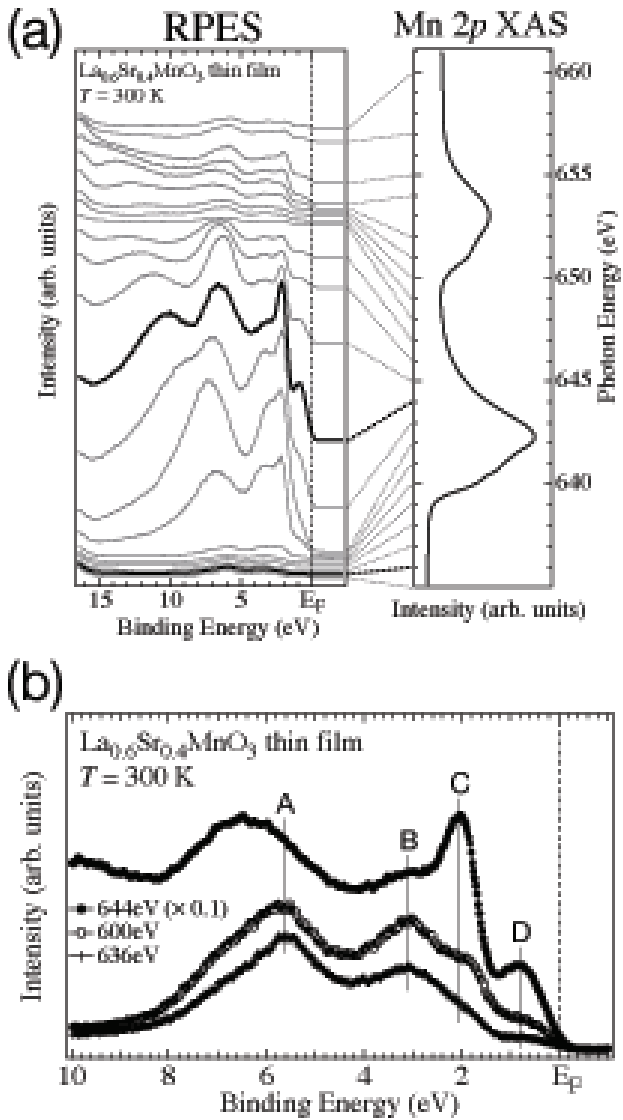}
\caption{\label{figure4}(a) (Left) {\it In situ} Mn 2$p$-3$d$ resonant PES spectra of La$_{0.6}$Sr$_{0.4}$MnO$_3$ thin films. Broken lines indicate the photon energy where the PES spectra were taken. Antiresonance (636 eV) and resonance (644 eV) spectra were shown with the black thick lines. (Right) Mn 2$p$ x-ray absorption spectrum of La$_{0.6}$Sr$_{0.4}$MnO$_3$ thin films. (b) Comparison between the antiresonance (crosses) and resonance (filled circles) PES spectra. Normal valence band PES spectrum taken with the photon energy of 600 eV (open circles) is also shown.}
\end{figure}

\subsection{Photoemission spectra on atomoically-controlled LSMO films}
Before proceeding to the compositional dependence of LSMO thin film, we first demonstrate the importance of PES measurements on a well-ordered surface. Figure~\ref{figure3} shows an {\it in situ} photoemission spectrum combined with an O 1$s$ XAS spectrum for the fabricated LSMO $x$ = 0.4 thin film. The cleanliness of the surface required for PES measurements is confirmed by the absence of a hump structure around the binding energy of 8 - 10 eV. The hump structure is typically seen in the PES spectra on contaminated transition-metal-oxides surfaces \cite{Saitoh1}. The contamination-free surface is also confirmed by the fact that the gcontaminationh signal at the higher binding energy of O 1$s$ core level was negligibly weak as shown in Fig.~\ref{figure6} later. In Fig.~\ref{figure3}, the valence band spectrum mainly consists of four structures as labeled $A$, $B$, $C$, and $D$. In order to check the character of these features, we carried out Mn 2$p$  - 3$d$ resonant PES (RPES) at various energies determined by the Mn 2$p$ XAS profile, as shown in Fig.~\ref{figure4} (a). Since the intensity of the structure $C$ and $D$ is resonantly enhanced at the photon energy around Mn 2$p$ - 3$d$ core threshold as shown in Fig.~\ref{figure4} (a) and (b), they originate from Mn 3$d$ states. In comparison with cluster model calculation \cite{Saitoh1}, the $C$ and $D$ structures are assigned as $t_{2g}$ and $e_g$ states, respectively. On the other hand, the two prominent structures ($A$ and $B$) have a dominant O 2$p$ character.

The structure near $E_F$ has not been clearly observed in the previous PES studies on scraped surfaces of LSMO crystal \cite{Chainani, Saitoh1} as well as these of other manganites \cite{JHPark_LCMO, Sekiyama_NSMO}, where the spectral intensity near $E_F$ is considerably suppressed. The strong enhancement of the near-$E_F$ feature in the present study is not due to the difference of the photoionization cross setion among the PES experiments, since the cross section ratio between the Mn 3$d$ and O 2$p$ orbitals at the photon energy of 600 eV is almost the same as that at the photon energy of 1253.6 eV (Mg K$\alpha$) \cite{Yeh_Lindau}.  The influence of the energy resolution is also ruled out, since the previous PES results could not be reproduced by broadening our spectrum with a Gaussian function to simulate the difference of the energy resolution in each experiment. In contrast, the O 1$s$ and Mn 2$p$ XAS spectra, which have much deeper probing depth than that of PES measurement, are in good agreement with the scraped one \cite{Abbate}. Therefore, we conclude that the suppression of the spectral intensity near $E_F$ may originate from surface disorder induced by scraping or fracturing, indicating the importance of {\it in situ} PES measurement on the well-ordered surface for revealing their intrinsic electronic structure.

\begin{figure}
\includegraphics[width=0.95\linewidth]{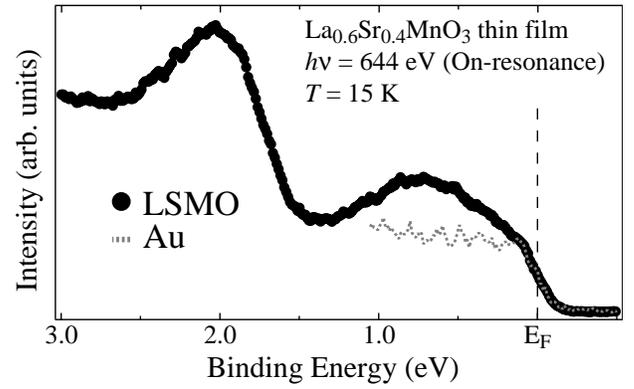}
\caption{\label{figure5}High-resolution resonant PES spectrum of La$_{0.6}$Sr$_{0.4}$MnO$_3$ thin films (filled circles) and photoemission spectrum of gold (broken line) near $E_F$ at 15 K.}
\end{figure}

The next crucial issue is whether the Fermi edge is clearly observed or not. In order to investigate the electronic structure near $E_F$ in more detail, we have measured the Mn 2$p$-3$d$ resonant PES spectra near $E_F$ with higher energy resolution at low temperature. Figure~\ref{figure5} shows the high-resolution Mn 2$p$ - 3$d$ RPES (HR-RPES) spectrum of the LSMO $x$ = 0.4 thin films. We find that the HR-RPES spectrum clearly exhibits a Fermi edge, reflecting the metallic ground states of LSMO $x$ = 0.4 thin films. The existence of a Fermi edge is more clearly seen by comparison with the spectrum of gold. This result is a sharp contrast to the previous PES results where the spectral weight near $E_F$ is considerably suppressed and consequently the Fermi edge is hardly seen \cite{Chainani, Saitoh1}. The suppression may originate from surface disorder induced by conventional surface preparation procedures (scraping or fracturing) in the conventional PES measurements, since the metallic state of LSMO results from the coherence of doped states which may be deeply disturbed by the disorder. In fact, the physical properties of manganites are considerably sensitive to the structural disorder \cite{CMR_Book}. These results strongly suggest the importance of {\it in situ} PES measurement on a well-ordered surface of transition metal oxides for revealing their intrinsic electronic structure.

\begin{figure}
\includegraphics[width=0.95\linewidth]{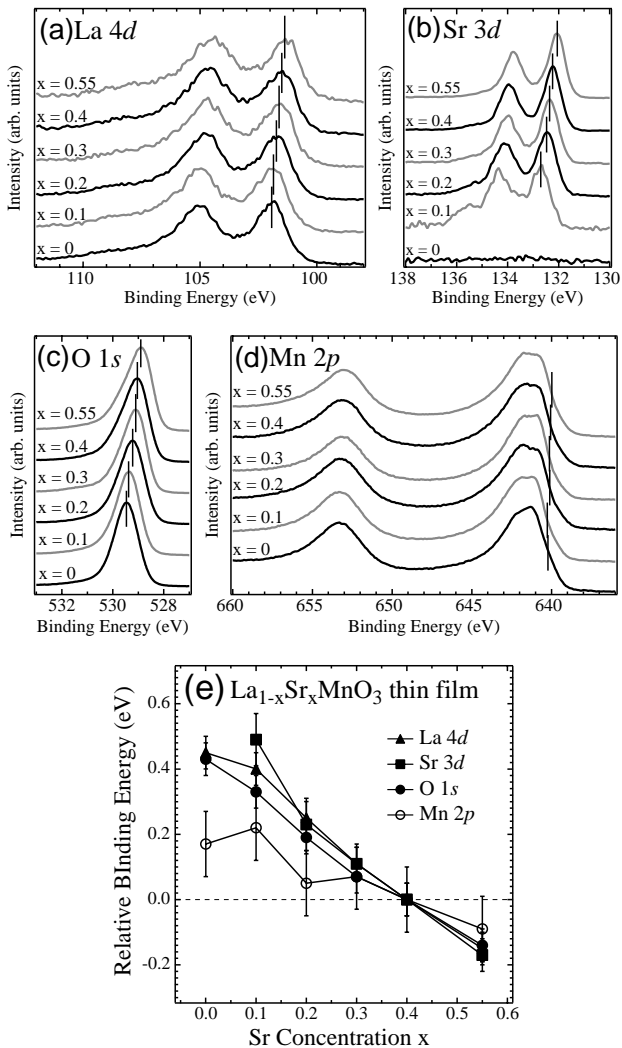}
\caption{\label{figure6}Composition dependence of (a) La 4$d$, (b) Sr 3$d$, (c) O 1$s$, and (d) Mn 2$p$ core-level PES spectra of LSMO thin films. (e) Composition dependence of peak positions of core-level PES spectra. We estimate the peak positions of La 4$d$, Sr 3$d$, and O 1$s$ core levels by fitted with a Gaussian function, and the peak positions of Mn 2$p$ core levels from the midpoint of the lower binding energy slope as indicated by vertical lines in Fig. 6 (a), (b), (c), and (d), respectively.}
\end{figure}

\subsection{Compositional dependence of core-level PES spectra}
Next, we discuss how the electronic structure changes as a function of hole doping.  Figure~\ref{figure6} shows Sr-concentration dependence of the core-level PES spectra obtained from well-ordered surfaces of LSMO thin films.  The peak positions of La 4$d$, Sr 3$d$, and O 1$s$ core levels were determined by a curve fitting as indicated by vertical lines in Figs.~\ref{figure6} (a), (b), and (c), respectively.  These core levels monotonically shift toward lower binding energy with increasing $x$.  The monotonic peak shift may reflect the chemical potential shift of LSMO with hole doping.  For Mn 2$p$ core levels, we have estimated the peak positions from the midpoint of a leading edge as indicated by vertical lines in Fig.~\ref{figure6} (d), since it is difficult to determine the peak positions of Mn 2$p_{3/2}$ core levels owing to the existence of Mn$^{3+}$ and Mn$^{4+}$ multiplets.  The estimation of energy shift from Mn 2$p_{1/2}$ peaks gives nearly the same results. The deviation of the peak shift for Mn 2$p$ core levels with respect to other core levels may be attributed to the increase in the Mn valence from Mn$^{3+}$ to Mn$^{4+}$ with hole doping, i.e., to the so-called chemical shift.  All core levels were found to shift by the same amount toward lower binding energy with increasing Sr concentration, reflecting a monotonic chemical potential shift with Sr concentration as summarized in Fig.~\ref{figure6} (e).  The monotonic core-level shift with hole-doping seems to be common phenomena in hole-doped manganite oxides \cite{Matsuno, JHPark_LCMO}.  The linear dependence of chemical potential shift as a function of $x$ is in contrast with the results on La$_{2-x}$Sr$_x$CuO$_4$ (LSCO) \cite{Fujimori_JESRP} and La$_{1-x}$Sr$_x$FeO$_3$ (LSFO) \cite{Wadati}, where the significant suppression of chemical potential shift has been observed in the under-doped region of LSCO and in the charge-ordered phase of LSFO, respectively. These phenomena have been attributed to the formation of charge stripe for LSCO and the charge disproportion for LSFO, which are a kind of gmicroscopic phase separationh.  Since electronic phase separation results in the pinning of the chemical potential, the monotonic peak shift indicates that the change in the electronic structure of LSMO with hole doping is well described by a rigid-band picture, suggesting the absence of phase separation on a microscopic scale \cite{Matsuno}.

\subsection{Compositional dependence of valence band PES spectra}
\begin{figure}
\includegraphics[width=0.95\linewidth]{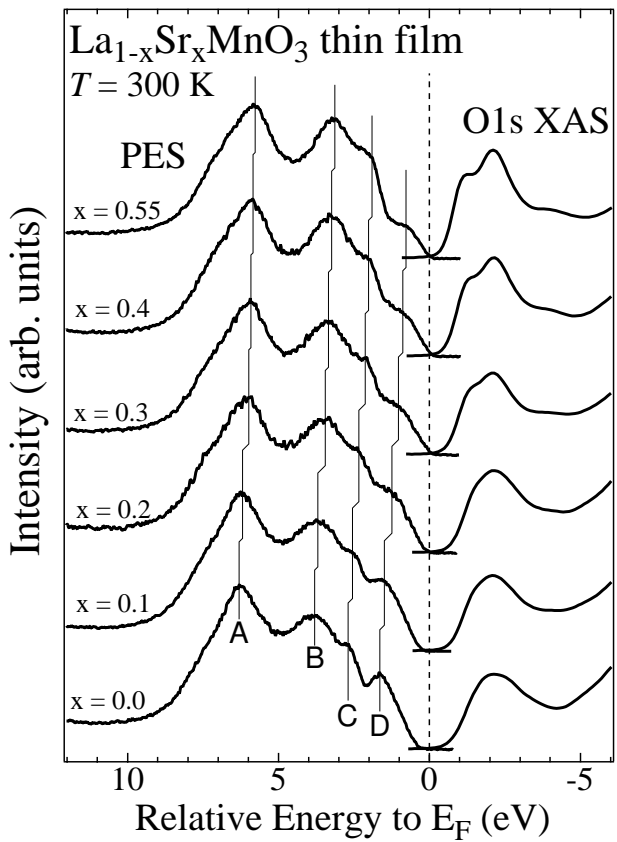}
\caption{\label{figure7}Composition dependence of valence band PES and O 1$s$ XAS spectra of LSMO thin films.  Peak positions indicated with vertical lines are determined by fitting of each spectrum with Gaussian components. The energies of XAS spectra were referenced to the optical gap of LaMnO$_3$ reported by Arima et al. \cite{Arima} and the $x$-dependent shift of O 1$s$ core-level peak in our PES measurements.}
\end{figure}

Figure~\ref{figure7} shows Sr-concentration dependence of the combined valence band PES spectra and O 1$s$ XAS spectra. Here, O 1$s$ XAS spectra for LaMnO$_3$ have been aligned so that the gap magnitude agrees with that obtained from the optical measurements, 1.1 eV \cite{Arima}, since the $E_F$ position for O 1$s$ XAS cannot be determined unambiguously from the O 1$s$ core-level PES and XAS data because of the unknown effect of the core-hole potential \cite{Saitoh1}. The Fermi levels of the XAS spectra for other compositions have been determined by combining the Fermi level position in the LaMnO$_3$ spectrum with the $x$-dependent shift of O 1$s$ core-level peak for sake of convenience.  In the valence band PES spectra for all compositions, we immediately notice that the features closest to $E_F$ are clearly observed in contrast to the previous PES results. Especially for LaMnO$_3$, the $e_g$-derived structure closest to $E_F$ is remarkably enhanced in comparison with the previous PES results \cite{Chainani, Saitoh1, JHPark_LCMO}.  The presence of such an $e_g$-derived feature in valence band spectra of LaMnO$_3$ is predicted by the recent band structure calculation considering the photoemission final state effect correctly \cite{Ravindran}. The overall features of the present PES spectrum of LaMnO$_3$ show good agreement with this calculation. In addition, we find that there is hardly any intensity in PES spectrum of LaMnO$_3$ at $E_F$, which is consistent with the insulating ground state of LaMnO$_3$. The absence of spectral weight within the band gap of LaMnO$_3$ suggests the influence of excess oxygen in LaMnO$_3$ is negligible.

With hole doping by substitution of Sr for La in LSMO, systematic change in the electronic structures has been observed. It was found that all structures of valence band PES spectra monotonically shift toward lower binding energy with increasing Sr concentration, which is the same as the core level spectra. On the other hand, the peak intensity of the feature $D$ systematically decreases with increasing the Sr concentration $x$, while the intensity of the other features at higher binding energies seems to be preserved. For compensating the loss of spectral weight, a shoulder structure near $E_F$ in O 1$s$ XAS spectra simultaneously increases, suggesting the spectral weight transfer across a gap or pseudogap (depression in spectral weight) at $E_F$.

\begin{figure}
\includegraphics[width=0.95\linewidth]{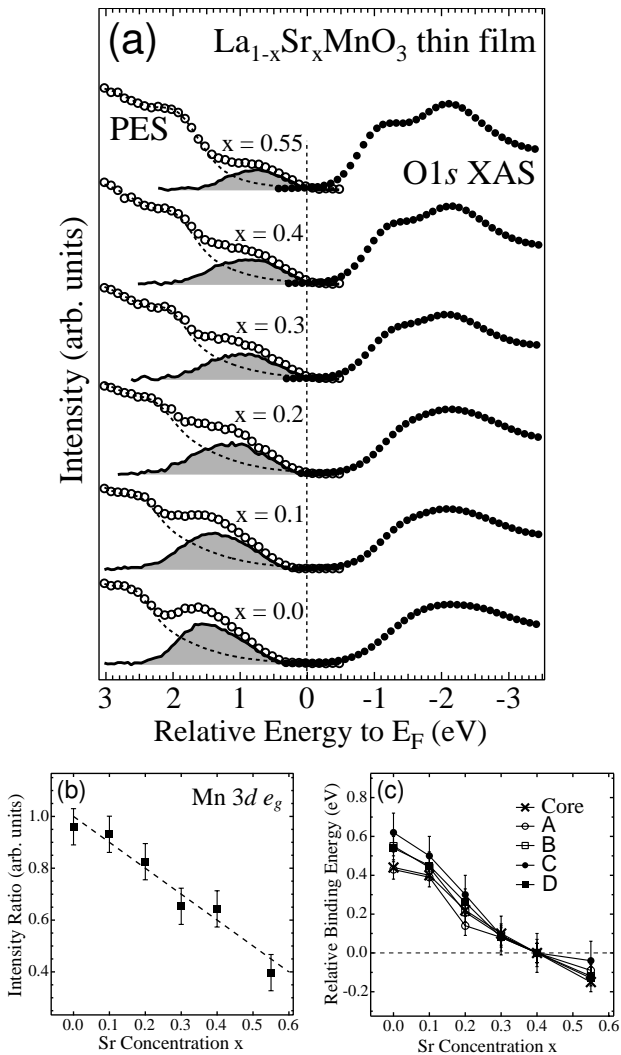}
\caption{\label{figure8}(a) The expansion graph of valence band PES spectra (open circles) and XAS spectra (filled circles) near $E_F$, and the extracted $e_g$ states (solid lines) obtained by subtracting the contribution from the other states (dotted lines) from the valence band PES spectra.  For detail, see text.  (b) Peak area of the extracted $e_g$ states plotted as a function of Sr concentration $x$.  A broken line shows the hole concentrations of these compounds.  (c) Composition dependence of peak positions of valence band PES spectra as well as core levels.}
\end{figure}

Figure~\ref{figure7} shows that features $A$, $B$ and $C$ shift in energy position but do not change significantly in intensity, while the $e_g$ state (feature $D$) shows a systematic change in intensity and peak position. In order to investigate the change in the $e_g$ state as a function of hole-doping level in more detail, we extract the $e_g$ states by subtracting the contribution from other states which are simulated by a linear combination of Gaussian functions and an integral background. This is under the assumption that features $A$, $B$ and $C$ effectively represent the "SrMnO$_3$ spectrum". The results are shown in Fig.~\ref{figure8} (a). It is clearly observed that the peak shift and the reduction of spectral weight occur in the $e_g$ states at the same time. Assuming the hole doped into the $e_g$ states mainly, the spectral weight should be proportional to the filing of $e_g$ states (1-$x$).  In Fig.~\ref{figure8} (b), we plot the relative spectral weight of the $e_g$ states as a function of hole concentraton, confirming a linear relation between the spectral weight and the filling of $e_g$ states.

The energy shifts in the valence band features are summarized in Fig.~\ref{figure8} (c), together with the average of the peak shifts in the La 4$d$, Sr 3$d$, and O 1$s$ core levels. As expected, these energy shifts are almost the same as each other, indicating the rigid-band behavior of underlying electronic structures of LSMO thin films. The observed monotonic shift of energy position of $e_g$ states is different from the previous PES studies on LSMO polycrystalline samples, where the energy shift of $e_g$ states got pinned in the metallic region of LSMO \cite{Saitoh1}, although a monotonic chemical potential shift for core levels was observed \cite{Matsuno, JHPark_LCMO}.  The discrepancy in energy shift among experiments may stem from the difficulty in determining the energy position accurately in the conventional PES measurements due to the considerable suppression of spectral weight for $e_g$ states.  Again, the reasonable spectral behavior with doping demonstrates that the high-quality spectra showing intrinsic spectral features in LSMO are successfully obtained from the present PES studies on well-ordered surfaces.

Finally, we discuss the redistribution of spectral weight near $E_F$. In the rigid band picture, it is expected that the peak intensity itself does not change unless the peak top crosses $E_F$. However, the peak intensity of $e_g$ states decreases linearly with hole doping as shown in Fig.~\ref{figure8} (a), together with simultaneous increase of intensity for shoulder structures near the leading edge of O 1$s$ XAS spectra. These results indicate that the gnon-rigid-band-likeh behavior occurs at the near-$E_F$ region, where a gap or pseudogap is opened at $E_F$ and consequently spectral weight is redistributed between the occupied states and unoccupied states with hole-doping \cite{Fujimori_JESRP}.  The systematic spectral weight transfer across $E_F$ is clearly seen in the combined PES and XAS spectra. This non-rigid-band behavior near $E_F$ is apparently in conflict with the monotonic chemical potential shift. We therefore conclude that the effect of hole-doping to the electronic structure of LSMO can be described in the framework of the rigid-band model in terms of energy positions of core-level and valence states, whereas the $e_g$ states also show non-rigid-band-like behavior in terms of spectral weight transfer from below to above $E_F$ across the gap or pseudogap at $E_F$.  

\section{CONCLUSION}
We have performed {\it in situ} synchrotron-radiation photoemission measurements on the well-ordered surface of LSMO thin films to investigate change in the electronic structure of LSMO thin films as a function of hole concentration $x$. The PES spectra for the laser MBE grown samples clearly show that the intensity of $e_g$ states closest to $E_F$ is significantly enhanced for well-ordered surfaces grown by laser MBE. Furthermore, high-resolution Mn 2$p$ - 3$d$ resonant PES spectra for metallic LSMO $x$ = 0.4 thin films exhibits the clear evidence of a Fermi cut-off, which is not clearly observed in the previous PES measurements.  These results clearly indicate the importance of {\it in situ} PES study on a well-ordered surface of transition metal oxides.  Combining the PES spectra and O 1$s$ XAS spectra on the well-ordered surface of LSMO thin films, we have successfully obtained a picture of how the electronic structure evolves as a function of hole concentration $x$.  The shift of chemical potential is proportional to $x$, indicating the rigid-band like behavior of underlying electronic structure of LSMO thin films.  In the PES spectra near $E_F$, the $e_g$-derived structure becomes weaker and moves toward $E_F$ as $x$ is increased. The linear relation between the spectral weight and the filling for $e_g$ states reveals that the holes are doped into the $e_g$ states. The pseudogap which is depression in spectral weight at $E_F$ exists for all metallic compositions. These results indicate that the simple rigid band model does not describe the electronic structure of LSMO and the spectral weight transfer occurs across $E_F$ in a non-rigid-band-like manner.

\begin{acknowledgments}
We thank Prof. A. Fujimori and Dr. A. Chainani for helpful discussions. This work has been done under Project 02S2-002 at the Institute of Material Structure Science at KEK.
\end{acknowledgments}


\begin{references}

\bibitem{MIT_Rev} M. Imada, A. Fujimori, and Y. Tokura, Rev. Mod. Phys. {\bf70}, 1039 (1998).

\bibitem{CMR_Book} {\it Colossal Magnetoresistive Oxides}, vol. 2 of {\it Advances in Condensed Matter Science}, edited by Y. Tokura (Gordon and Breach, Amsterdam, 2000).

\bibitem{Urushibara} A. Urushibara, Y. Moritomo, T. Arima, A. Asamitsu, G. Kido, and Y. Tokura, Phys. Rev. B {\bf51}, 14103 (1995).

\bibitem{JHPark} J. -H. Park, E. Vescovo, H. J. Kim, C. Kwon, R. Ramesh, and T. Venkatesan, Nature {\bf392}, 794 (1998).

\bibitem{Fujishiro} H. Fujishiro, M. Ikebe, and Y. Konno, J. Phys. Soc. Jpn. {\bf67}, 1799 (1998).

\bibitem{Chainani} A. Chainani, M. Mathew, and D. D. Sarma, Phys. Rev. B {\bf47}, 15397 (1993).

\bibitem{Saitoh1} T. Saitoh, A. E. Bocquet, T. Mizokawa, H. Namatame, A. Fujimori, M. Abbate, Y. Takeda, and M. Takano, Phys. Rev. B {\bf51}, 13942 (1995).

\bibitem{Sarma} D. D. Sarma, N. Shanthi, S. R. Krishnakumar, T. Saitoh, T. Mizokawa, A. Sekiyama, K. Kobayashi, A. Fujimori, E. Weschke, R. Meier, G. Kaindl, Y. Takeda, and M. Takano, Phys. Rev. B {\bf53}, 6873 (1996).

\bibitem{Saitoh2} T. Saitoh, A. Sekiyama, K. Kobayashi, T. Mizokawa, A. Fujimori, D. D. Sarma, Y. Takeda, and M. Takano, Phys. Rev. B {\bf56}, 8836 (1997).

\bibitem{Matsuno} J. Matsuno, A. Fujimori, Y. Takeda, and M. Takano, Europhys. Lett. {\bf59}, 252 (2002).

\bibitem{Sekiyama_Nature} A. Sekiyama, T. Iwasaki, K. Matsuda, Y. Saitoh, Y. Onuki, and S. Suga, Nature {\bf403}, 396 (2000).

\bibitem{Joynt} R. Joynt, Science {\bf284}, 777 (1999).

\bibitem{Koinuma}H. Koinuma, N. Kanda, J. Nishino, A. Ohtomo, H. Kubota, M. Kawasaki, and M. Yoshimoto, Appl. Surf. Sci. {\bf109/110}, 514 (1997).

\bibitem{Horiba_RSI}K. Horiba, H. Ohguchi, H. Kumigashira, M. Oshima, K. Ono, N. Nakagawa, M. Lippmaa, M. Kawasaki, and H. Koinuma, Rev. Sci. Instrum. {\bf74}, 3406 (2003).

\bibitem{BL2C}M. Watanabe, A. Toyoshima, Y. Azuma, T. Hayaishi, Y. Yan, and A. Yagishita, Proc. SPIE {\bf3150}, 58 (1997).

\bibitem{Kawasaki}M. Kawasaki, K. Takahashi, T. Maeda, R. Tsuchiya, M. Shinohara, T. Yonezawa, O. Ishihara, M. Yoshimoto, and H. Koinuma, Science {\bf266}, 1540 (1994).

\bibitem{Kumi_APL}H. Kumigashira, K. Horiba, H. Ohguchi, K. Ono, M. Oshima, N. Nakagawa, M. Lippmaa, M. Kawasaki, and H. Koinuma, Appl. Phys. Lett. {\bf82}, 3430 (2003).

\bibitem{Izumi_APL} M. Izumi, Y. Konishi, T. Nishihara, S. Hayashi, M. Shinohara, M. Kawasaki, and Y. Tokura, Appl. Phys. Lett. {\bf73}, 2497 (1998).

\bibitem{Izumi_PRB} M. Izumi, T. Manako, Y. Konishi, M. Kawasaki, and Y. Tokura, Phys. Rev. B {\bf61}, 12187 (2000).

\bibitem{Fukumura} T. Fukumura, M. Ohtani, M. Kawasaki, Y. Okimoto, Y. Tokura, and H. Koinuma, Appl. Phys. Lett. {\bf77}, 3426 (2000).

\bibitem{JHPark_LCMO}J. -H. Park, C. T. Chen, S-W. Cheong, W. Bao, G. Meigs, V. Chakarian, and Y. U. Idzerda, Phys. Rev. Lett. {\bf76}, 4215 (1997).

\bibitem{Sekiyama_NSMO}A. Sekiyama, S. Suga, M. Fujikawa, S. Imada, T. Iwasaki, K. Matsuda, T. Matsushita, K. V. Kaznacheyev, A. Fujimori, H. Kuwahara, and Y. Tokura, Phys. Rev. B {\bf59}, 15528 (1999).

\bibitem{Yeh_Lindau}J. J. Yeh and I. Lindau, At. Data Nucl. Data Tables {\bf32}, 1 (1985).

\bibitem{Abbate}M. Abbate, F. M. F. de Groot, J. C. Fuggle, A. Fujimori, O. Strebel, F. Lopez, M. Domke, G. Kaindl, G. A. Sawatzky, M. Takano, Y. Takeda, H. Eisaki, and S. Uchida, Phys. Rev. B {\bf46}, 4511 (1992).

\bibitem{Fujimori_JESRP}A. Fujimori, A. Ino, J. Matsuno, T. Yoshida, K. Tanaka and T. Mizokawa, J. Electron Spectrosc. Relat. Phenom. {\bf124}, 127 (2002).

\bibitem{Wadati}H. Wadati, D. Kobayashi, H. Kumigashira, K. Okazaki, T. Mizokawa, A. Fujimori, K. Horiba, M. Oshima, N. Hamada, M. Lippmaa, M. Kawasaki, and H. Koinuma, cond-mat/0404435.

\bibitem{Arima}T. Arima, Y. Tokura, and J. B. Torrance, Phys. Rev. B {\bf48}, 17006 (1993).

\bibitem{Ravindran}P. Ravindran, A. Kjekshus, H. Fjellv{\aa}g, A. Delin, and O. Eriksson, Phys. Rev. B {\bf65}, 064445 (2002).

\end{references}
\end{document}